\documentclass[aps,prl,twocolumn,superscriptaddress]{revtex4-2}
\usepackage{graphicx}
\usepackage{latexsym}
\usepackage{amssymb}
\usepackage{amsmath}
\usepackage{amsfonts}
\usepackage{upgreek}
\usepackage{subfigure}
\usepackage{enumerate}
\usepackage{bm}
\usepackage{nicefrac}
\usepackage{bbold}
\usepackage[version=3]{mhchem}
\usepackage{verbatim}
\usepackage[unicode=true,
 bookmarks=false,
 breaklinks=false,pdfborder={0 0 1},backref=false,colorlinks=true]
 {hyperref}
\hypersetup{
 linkcolor=[rgb]{0,0,1},citecolor=[rgb]{0,0,1},urlcolor=[rgb]{0,0,1}}
\usepackage{multirow}
\usepackage{color}
\usepackage{comment}
\usepackage{xcolor}
\usepackage{soul}
\usepackage{tikz}

\usepackage{braket}

\begin{document}

\title{Superconductivity in the pressurized nickelate \ce{La3 Ni2 O7} \\ in the vicinity of a BEC-BCS crossover}


\author{Henning Schl\"omer}
\email{H.Schloemer@physik.uni-muenchen.de}
\affiliation{Department of Physics and Arnold Sommerfeld Center for Theoretical Physics (ASC), Ludwig-Maximilians-Universit\"at M\"unchen, Theresienstr. 37, M\"unchen D-80333, Germany}
\affiliation{Munich Center for Quantum Science and Technology (MCQST), Schellingstr. 4, D-80799 M\"unchen, Germany}
\author{Ulrich Schollw\"ock}
\affiliation{Department of Physics and Arnold Sommerfeld Center for Theoretical Physics (ASC), Ludwig-Maximilians-Universit\"at M\"unchen, Theresienstr. 37, M\"unchen D-80333, Germany}
\affiliation{Munich Center for Quantum Science and Technology (MCQST), Schellingstr. 4, D-80799 M\"unchen, Germany}
\author{Fabian Grusdt}
\affiliation{Department of Physics and Arnold Sommerfeld Center for Theoretical Physics (ASC), Ludwig-Maximilians-Universit\"at M\"unchen, Theresienstr. 37, M\"unchen D-80333, Germany}
\affiliation{Munich Center for Quantum Science and Technology (MCQST), Schellingstr. 4, D-80799 M\"unchen, Germany}
\author{Annabelle Bohrdt}
\affiliation{Munich Center for Quantum Science and Technology (MCQST), Schellingstr. 4, D-80799 M\"unchen, Germany}
\affiliation{Institut für Theoretische Physik, Universität Regensburg, D-93035 Regensburg, Germany}

\date{\today}
\begin{abstract}
Ever since the discovery of high-temperature superconductivity in cuprates, gaining microscopic insights into the nature of pairing in strongly correlated, repulsively interacting fermionic systems has remained one of the greatest challenges in modern condensed matter physics. Following recent experiments reporting superconductivity in the bilayer nickelate \ce{La3 Ni2 O7} (LNO) with remarkably high critical temperatures of $T_c = 80$ K~\cite{Sun2023}, it has been argued that the low-energy physics of LNO can be described by the strongly correlated, mixed-dimensional bilayer $t-J$ model~\cite{wu2023charge, lu2023, oh2023type, qu2023, jiang2023high}. Here we investigate this bilayer system and utilize density matrix renormalization group techniques to establish a thorough understanding of the model and the magnetically induced pairing through comparison to the perturbative limit of dominating inter-layer spin couplings. In particular, this allows us to explain appearing finite-size effects, firmly establishing the existence of long-range superconducting order in the thermodynamic limit that is described by the XXZ universality class of hard-core bosons constituted by $s$-wave singlet pairs. As the effective model in the perturbative limit is known to show linear resistivity above the superconducting transition temperature, we propose a pair-based interpretation of the extended strange metal phase observed in LNO. By analyzing binding energies, we predict a BEC-BCS crossover as a function of the Hamiltonian parameters, whereas LNO is anticipated to lie on the BCS side in vicinity of the transition. We find large binding energies of the order of the inter-layer coupling that suggest strikingly high critical temperatures of the Berezinskii-Kosterlitz-Thouless transition, raising the question whether (mixD) bilayer superconductors possibly facilitate critical temperatures above room temperature.
\end{abstract}
\maketitle

\textbf{Introduction.---}  Although the discovery of high-temperature superconductivity in cuprates dates back more than three decades~\cite{Bednorz1986, Lee2006, Anderson1987}, fully understanding their enigmatic pairing mechanism remains an unsolved and long-sought problem in contemporary condensed matter physics. In particular, detailed microscopic insights into the relevant physics are necessary to open the path towards a targeted design of novel materials, possibly with high critical temperatures at ambient conditions~\cite{Berg2008, Lee2018, bohrdt2021strong, Hirthe2022}. 

\begin{figure*}
\centering
\includegraphics[width=0.96\textwidth]{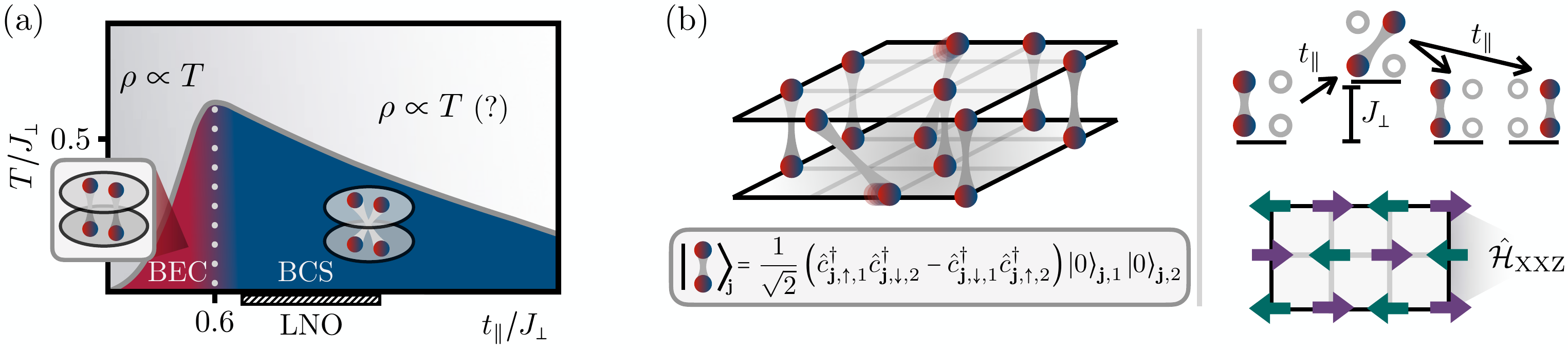}
\caption{\textbf{Schematic phase diagram and effective model.} (a) Schematic phase diagram of the mixD $t_{\parallel}-J_{\perp}-J_{\parallel}$ model, Eq.~\eqref{eq:Hbl}, at doping $\delta = 50\%$ relevant to LNO. In the limit of dominating inter-layer magnetic interactions, a BEC-type superfluid of tightly bound pairs is realized. When the average sizes of pairs become larger, spatially extended pairs form a BCS-like superconducting state. Binding energies and estimated critical temperatures in vicinity of the crossover are of the order of the magnetic coupling $J_{\perp}$. In the BEC regime, the model shows linear in $T$ resistivity ($\rho \propto T$) above the superconducting phase, which may extend to larger values of $t_{\parallel}/J_{\perp}$ above the BCS regime. The relevant parameter regime for LNO depending on the strength of the on-site repulsion is shown by the black hatched area, where $t_{\parallel}/J_{\perp} \sim 0.7 - 1.5$ (which may however be renormalized when taking into account multi-band effects). (b) In the limit $J_{\perp} \gg t_{\parallel}, J_{\parallel}$, the bilayer mixD $t_{\parallel}-J_{\perp}-J_{\parallel}$ model, Eq.~\eqref{eq:Hbl}, reduces to a model of hopping singlets (left panel). Singlets hop on the bilayer structure via second order processes (right panel, upper illustration), leading to a single layer interacting hard-core bosonic system, Eq.~\eqref{eq:Hhcb}, in the perturbative limit. A further mapping to a spin system yields an effective 2D $\text{XXZ}$ model, Eq.~\eqref{eq:XXZ}, where spin-spin correlations in the $xy$-plane map to coherent pair-pair correlations in the bilayer system (right panel, lower illustration).}
\label{fig:PT}
\end{figure*}

Very recently, the Ruddlesen-Popper bilayer perovskite nickelate \ce{La3 Ni2 O7} (LNO) has joined the family of bulk superconductors above the boiling point of liquid nitrogen, with extraordinarily high critical temperatures of $T_c = 80$ K at applied pressures above $14$ GPa~\cite{Sun2023, zhang2023_zeroR, hou2023emergence}. Density functional theory (DFT) calculations suggest that the active degrees of freedom near the Fermi energy in the layered LNO structure are given by the $3 d_{x^2 - y^2}$ and $3 d_{z^2}$ $\ce{Ni}$ orbitals~\cite{Pardo2011, Luo2023, zhang2023, sakakibara2023possible, gu2023, christiansson2023correlated, cao2023flat}, whereby the four $3d$ orbitals in each unit cell (two in each layer) share three electrons. The $3d$ character of the electronic structure together with the absence of perfect nesting in the non-interacting model indicates the necessity of strong coupling approaches for an accurate description of LNO~\cite{yang2023minimal}, in line with recent experiments suggesting the vicinity of LNO to a Mott transition~\cite{Liu2023}. 

Starting in the limit of strong on-site repulsion, two of the three electrons in each unit cell fill the $d_{z^2}$ orbitals due to lower on-site energies of the $3d_{z^2}$ compared to the $3d_{x^2 - y^2}$ states. This results in a half filled, Mott insulating $d_{z^2}$ band, while the $d_{x^2 - y^2}$ orbitals at quarter filling constitute an itinerant, conducting band~\cite{shen2023}. Hybridization of the $3 d_{z^2}$-$\ce{Ni}$ and apical $2p_z$-$\ce{O}$ orbitals has been demonstrated to mediate strong inter-layer couplings between the $d_{z^2}$ orbitals of the two $\ce{Ni}$ layers within each unit cell at high pressures, where the $\ce{Ni}$-$\ce{O}$-$\ce{Ni}$ bonding angles are aligned at an angle of $180^{\circ}$ (the crystal structure of LNO experiences a structural transition from the \textit{Amam} to the higher-symmetry \textit{Fmmm} space group at pressures $\sim 10$ GPa)~\cite{Sun2023}.

The inter-layer superexchange between the insulating $d_{z^2}$ spins has been argued to be elevated to the $d_{x^2-y^2}$ orbitals by strong intra-atomic Hund's couplings~\cite{cao2023flat}, whereby the formation of a spin-triplet between the two active orbitals at each site is favored. Integrating out the $d_{z^2}$ degrees of freedom yields a minimal, single-band, mixed-dimensional (mixD) bilayer $t_{\parallel}-J_{\perp}-J_{\parallel}$ model for describing the essential low-energy physics of LNO~\cite{lu2023, qu2023, oh2023type, jiang2023high},
\begin{equation}
\begin{aligned}
    \hat{\mathcal{H}}_{\text{BL}} = -t_{\parallel} &\sum_{ \braket{\mathbf{i}, \mathbf{j}}, \sigma, \alpha} \hat{\mathcal{P}} \big(\hat{c}_{\mathbf{i}, \sigma, \alpha}^{\dagger} \hat{c}_{\mathbf{j}, \sigma, \alpha}^{\vphantom\dagger} + \text{h.c.} \big)\hat{\mathcal{P}} \,  \\ & + J_{\parallel} \sum_{\braket{\mathbf{i}, \mathbf{j}}, \alpha} \left( \hat{\mathbf{S}}_{\mathbf{i},\alpha} \cdot \hat{\mathbf{S}}_{\mathbf{j},\alpha} - \frac{\hat{n}_{\mathbf{i},\alpha}\hat{n}_{\mathbf{j},\alpha}}{4} \right)  \\ &  \quad \, \, \, + J_{\perp} \sum_{\mathbf{i}} \left( \hat{\mathbf{S}}_{\mathbf{i},1} \cdot \hat{\mathbf{S}}_{\mathbf{i},2} - \frac{\hat{n}_{\mathbf{i},1}\hat{n}_{\mathbf{i},2}}{4} \right).
\end{aligned}
\label{eq:Hbl}
\end{equation}
Here, $\hat{c}_{\mathbf{i}, \sigma, \alpha}^{(\dagger)}$, $\hat{n}_{\mathbf{i}, \alpha}$ and $\hat{\mathbf{S}}_{\mathbf{i}, \alpha}$ are fermionic annihilation (creation), particle density, and spin operators on site $\mathbf{i}$ and layer $\alpha = 1,2$, respectively; $\braket{\mathbf{i}, \mathbf{j}}$ denotes nearest neighbor (NN) sites on the two-dimensional (2D) square lattice, and $\hat{\mathcal{P}}$ is the Gutzwiller operator projecting out states with double occupancy. To describe LNO, the quarter filled $d_{x^2 - y^2}$ band corresponds to a doping level of $\delta = 0.5$ in the mixD $t_{\parallel}-J_{\perp}-J_{\parallel}$ model compared to the half-filled state with one particle per site. 

The Hamiltonian Eq.~\eqref{eq:Hbl} and related, multi-band models have been studied in particular with regard to pairing order using matrix product~\cite{qu2023, kaneko2023, shen2023, Chen2024orbital} and mean field~\cite{oh2023type, lu2023, lu2023superconductivity} methods, supporting the appearance of inter-layer $s$-wave superconductivity. We note that while a multi-band description that includes both $d_{x^2 - y^2}$ and $d_{z^2}$ orbitals provides a more accurate description of LNO (e.g. capturing self-doping effects between the two active orbitals~\cite{cao2023flat, oh2023type}), studying the single-band model may allow for a detailed understanding of the essential physics governing superconductivity in mixD bilayer systems. In particular, though it is commonly agreed upon that inter-layer magnetic interactions provide the pairing glue for superconductivity in the bilayer model~\cite{bohrdt2021strong, Hirthe2022, qu2023}, predicting and understanding the structure of the ground state in the simplified single-band model may be particularly useful for engineering materials with high critical temperatures.

Here, we use matrix product methods to study pair-pair correlations as well as binding energies in the Hamiltonian Eq.~\eqref{eq:Hbl} on finite width bilayer geometries. By comparison to the limit of strong inter-layer spin-spin interactions, where the model can be mapped to a spin-$1/2$ XXZ model, we gain a detailed understanding of the mixD $t_{\parallel}-J_{\perp}-J_{\parallel}$ model even away from this perturbative limit. In particular, this allows us to understand appearing finite-size effects and the influence of the various coupling parameters on the long-range pairing order, where the latter might permit to realize a certain tunability of experimental probes to favorable situations. We note that multi-band models taking into account Hund's coupling in a rigorous manner have been shown to reduce to the single-band Hamiltonian Eq.~\eqref{eq:Hbl} in the limit $J_{\perp}\gg t_{\parallel}$, however with weaker effective interlayer couplings~\cite{oh2023type, yang2023strong}. This suggests that, while energy scales may be renormalized, the single-band mixD $t$-$J$ model captures the essential low-energy physics of more accurate multi-band models. 

Through the computation of binding energies, we anticipate the emergence of a crossover from a Bose-Einstein condensate (BEC) to Bardeen-Cooper-Schrieffer (BCS) state in the mixD bilayer model as a function of $t_{\parallel}/J_{\perp}$, see Fig.~\ref{fig:PT}~(a), characterized by extended, overlapping pairs. We estimate critical temperatures of the superfluid transition to be of the order of the magnetic coupling, hence possibly facilitating superconductivity at temperatures beyond room temperature in mixD bilayer systems. In addition, as the effective model of tightly-bound pairs in the limit of strong spin couplings $J_{\perp} \gg t_{\parallel}, J_{\parallel}$ yields a linear resistivity as a function of temperature above the superconducting phase, we speculate that the resistivity in the bilayer model in vicinity to the crossover is governed by the conduction of pairs. Our results are summarized in the schematic phase diagram in Fig.~\ref{fig:PT}~(a).

Our work presents the perturbative limit of dominating inter-layer spin couplings as an important case study that allows for an understanding of qualitative physical features even away from this limit. This is in stark contrast to the Fermi-Hubbard model -- believed to capture the essential physics of cuprate superconductors -- where such controlled perturbative limits are absent, and large scale numerical simulations are necessary to resolve the small energy differences of competing phases~\cite{LeBlanc2015, Zheng2017, Jiang2019_science, Qin_absence_SC, MultiMess2021, Jiang_Kivelson, Arovas2022_rev, Qin_perspective, xu2023coexistence}. In fact, in the Fermi-Hubbard model, though tremendous progress has been made in recent years, its phase diagram (and in particular its applicability to capture the phases appearing in cuprate superconductors) is still under active debate. 

Though the single-band model, Eq.~\eqref{eq:Hbl}, is not believed to quantitatively describe e.g. transition temperatures of pressurized bilayer nickelates, establishing a microscopic understanding of simplified models by fully taking into account their correlation structure is an important step towards developing a theory of bilayer superconductors. Our calculations suggest high critical temperatures of the single-band model, which facilitates the preparation of a state with (quasi) long-range superconducting order in ultracold atom experiments with currently realistic temperatures. This, in turn, may allow for a systematic exploration of novel materials using analog quantum simulation platforms. 

\textbf{Perturbative limit.---} In the case of dominating spin couplings $J_{\perp} \gg t_{\parallel}, J_{\parallel}$, the fermions pair into tightly bound inter-layer singlets, where breaking apart a singlet is associated with energy cost $J_{\perp}$. In this limit, the low-energy physics of Eq.~\eqref{eq:Hbl} is described by the restricted local basis consisting of empty sites on site $\mathbf{j}$ in both layers, $\ket{0}_{\mathbf{j}} = \ket{0}_{\mathbf{j},1} \ket{0}_{\mathbf{j},2}$ (a chargon-chargon pair), as well as paired singlets, $\ket{1}_{\mathbf{j}} = \hat{b}^{\dagger}_{\mathbf{j}} \ket{0}_{\mathbf{j}}$, where the (hard-core) bosonic operator $\hat{b}^{\dagger}$ creates an inter-layer spin singlet, $\hat{b}^{\dagger}_{\mathbf{j}} \ket{0}_{\mathbf{j}} = \frac{1}{\sqrt{2}}\left( \hat{c}^{\dagger}_{\mathbf{j},\uparrow, 1} \hat{c}^{\dagger}_{\mathbf{j},\downarrow,2} - \hat{c}^{\dagger}_{\mathbf{j},\downarrow,1} \hat{c}^{\dagger}_{\mathbf{j},\uparrow,2} \right) \ket{0}_{\textbf{j}}$. By considering virtual processes to spinon-chargon states $c^{\dagger}_{\mathbf{j}, \sigma, \alpha} \ket{0}_{\mathbf{j}}$ in second order perturbation theory, and restricting the effective Hamiltonian to the low-energy subspace, the mixD bilayer model reduces to an interacting hard-core bosonic system in a single 2D plane illustrated in Fig.~\ref{fig:PT}~(b), as shown in~\cite{Bohrdt2020} (see also~\cite{Lange2023_1, Lange2023_2}),
\begin{equation}
\begin{aligned}
    \hat{\mathcal{H}}_{\text{HCB}} = -\frac{K}{2} &\sum_{ \braket{\mathbf{i}, \mathbf{j}}} \hat{\mathcal{P}} \big(\hat{b}_{\mathbf{i}}^{\dagger} \hat{b}_{\mathbf{j}}^{\vphantom\dagger} + \text{h.c.} \big)\hat{\mathcal{P}} - J_{\perp} \sum_{\mathbf{i}} \hat{b}_{\mathbf{i}}^{\dagger} \hat{b}_{\mathbf{i}}^{\vphantom\dagger}  \\ & + K \sum_{\braket{\mathbf{i}, \mathbf{j}}} \left( \Delta \hat{b}_{\mathbf{i}}^{\dagger} \hat{b}_{\mathbf{i}}^{\vphantom\dagger} \hat{b}_{\mathbf{j}}^{\dagger} \hat{b}_{\mathbf{j}}^{\vphantom\dagger} - \frac{\hat{b}_{\mathbf{i}}^{\dagger} \hat{b}_{\mathbf{i}}^{\vphantom\dagger}}{2} - \frac{\hat{b}_{\mathbf{j}}^{\dagger} \hat{b}_{\mathbf{j}}^{\vphantom\dagger}}{2}  \right),
\end{aligned}
\label{eq:Hhcb}
\end{equation}
where $K = 4 t_{\parallel}^2/J_{\perp}$ and $\Delta = 1-J_{\parallel}/2K$. The density-density term of the in-plane Heisenberg interactions in Eq.~\eqref{eq:Hbl} leads to the appearance of an anisotropy $\Delta < 1$. The effective boson model Eq.~\eqref{eq:Hhcb}, in turn, can be mapped to a 2D XXZ spin system~\cite{sachdev_2023_qpm},
\begin{equation}
\begin{aligned}
    \hat{\mathcal{H}}_{\text{XXZ}} = K &\sum_{\braket{\mathbf{i}, \mathbf{j}}} \left( \hat{J}^x_{\mathbf{i}} \hat{J}^x_{\mathbf{j}} + \hat{J}^y_{\mathbf{i}} \hat{J}^y_{\mathbf{j}} + \Delta \hat{J}^z_{\mathbf{i}} \hat{J}^z_{\mathbf{j}} \right) \\ & - J_{\perp} \sum_{\mathbf{i}} \hat{J}_{\mathbf{i}}^{z} - \frac{J_{\parallel}}{4} \sum_{\braket{\mathbf{i}, \mathbf{j}}} \left(\hat{J}^z_{\mathbf{i}} +  \hat{J}^z_{\mathbf{j}} \right) ,
\end{aligned}
\label{eq:XXZ}
\end{equation}
where a unitary transformation has been applied to make all coefficients positive and trivial constant terms have been dropped.  $\hat{J}^{\mu}_{\mathbf{i}}$, $\mu = x,y,z$ are spin-$1/2$ operators -- not to be confused with the spin operators $\hat{S}^{\mu}_{\mathbf{i}}$ of the fermionic bilayer Hamiltonian, Eq.~\eqref{eq:Hbl}. The magnetization of the spin model maps to the filling $\delta$ of the bilayer model as $m = \delta - 1/2$. The last term in Eq.~\eqref{eq:XXZ} is constant in periodic systems, however induces non-trivial effects for open boundaries~\cite{supp_mat}. For $J_{\parallel} = 0$, Eq.~\eqref{eq:XXZ} reduces to the Heisenberg model with an emerging $\rm{SU(2)}$ symmetry. 

In the perturbative regime, the bilayer system is hence a \textit{bona fide} superconductor, featuring long-range pairing order in the ground state that translates to long-range antiferromagnetic order in the $xy$-plane of the XXZ model, see the lower right panel in Fig.~\ref{fig:PT}~(b). The controlled connection to the XXZ model in the perturbative limit will prove to be useful in the following analysis of the appearing phases in the mixD bilayer model.

\begin{figure}
\centering
\includegraphics[width=\columnwidth]{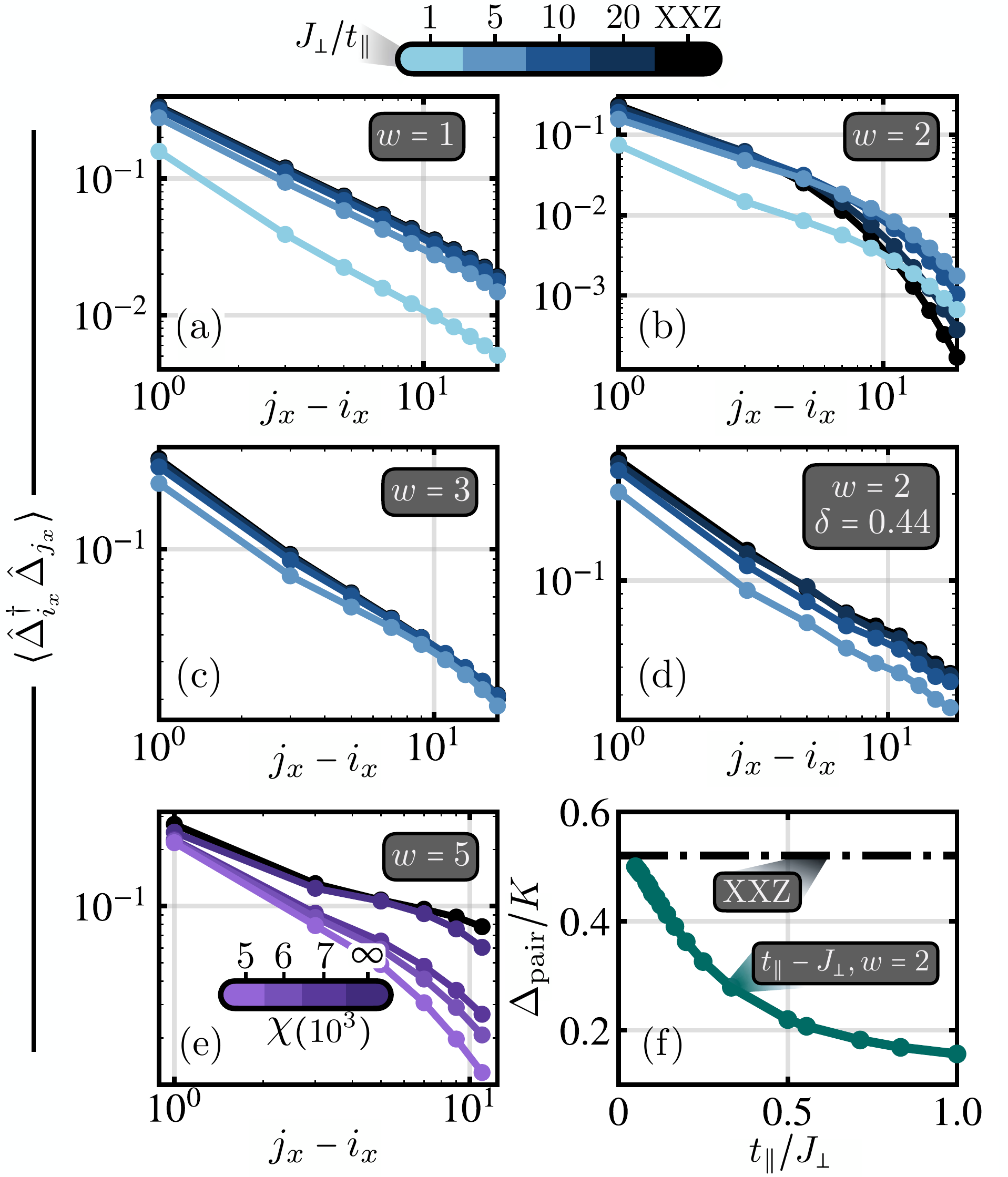}
\caption{\textbf{Pair correlations.} Pair-pair correlation function $\braket{\hat{\Delta}^{\dagger}_{\mathbf{i}} \hat{\Delta}_{\mathbf{j}}^{\vphantom\dagger}}$ in the $t_{\parallel}-J_{\perp}$ model ($J_{\parallel} = 0$) for varying $J_{\perp}/t_{\parallel}$ and width $w$; $\delta = 0.5$ is used if not indicated differently. For $w=1$ (a) and $w=3$ (c), correlations show algebraic signals, with increasing magnitudes for growing $J_{\perp}/t_{\parallel}$ while the decay exponents stay almost constant. Pair-pair correlations converge towards spin-spin correlations $\braket{\hat{J}^+_{\mathbf{i}} \hat{J}^-_{\mathbf{j}}}$ of the XXZ model in the perturbative limit (black data). For $w=5$, an extrapolation to large bond dimension for $J_{\perp}/t_{\parallel} = 20$ is shown, matching the prediction from the XXZ model. For $w=2$ at $\delta = 0.5$ (b), the decay of $\braket{\hat{\Delta}^{\dagger}_{\mathbf{i}} \hat{\Delta}_{\mathbf{j}}^{\vphantom\dagger}}$ is exponential for all values of $J_{\perp}/t_{\parallel}$, with decreasing correlation lengths for increasing $J_{\perp}/t_{\parallel}$. This is explained by a finite pair charge gap $\Delta_{\text{pair}}$ that corresponds to the spin gap of the SU(2) symmetric Heisenberg model in the perturbative limit (f). Away from $\delta = 0.5$ and for $w=2$ (d), correlations decay algebraically for all values of $J_{\perp}/t_{\parallel}$, as expected from the Heisenberg model at finite magnetization. We choose reference sites $\mathbf{i} = [i_x=10, i_y = 1]$ ($l = 32$) for $w=1,2$, $\mathbf{i} = [i_x=4, i_y = 2]$ ($l=24$) for $w=3$ and $\mathbf{i} = [i_x=2, i_y = 3]$ ($l=16$) for $w=5$.}
\label{fig:ppc}
\end{figure}

\textbf{Pair correlations and finite-size effects.---} We simulate the bilayer $t_{\parallel}-J_{\perp}-J_{\parallel}$ system, Eq.~\eqref{eq:Hbl}, in the ground state using the density matrix renormalization group~\cite{Schollwoeck_DMRG, SchollwoeckDMRG2, WhiteDMRG, hubig:_syten_toolk, hubig17:_symmet_protec_tensor_networ} for various parameters $J_{\perp}/t_{\parallel}$ at $J_{\parallel} = 0$ and doping $\delta = 0.5$. We focus on systems of size $l \times w \times 2$, where $w$ and $l$ are the width and length of each layer in the bilayer system, respectively. We implement separate $\rm{U(1)}$ symmetries in each layer and conserve the total magnetization, such that the symmetry of the system is given by $\rm{U(1)}^{\alpha = 1} \otimes \rm{U(1)}^{\alpha = 2}  \otimes \rm{U(1)}^{S^z_{\text{tot}}}$. We use bond dimensions up to $\chi = 6000$ and carefully ensure that our results are converged~\cite{supp_mat}.

Fig.~\ref{fig:ppc} shows coherent pair-pair correlations $\braket{\hat{\Delta}^{\dagger}_{\mathbf{i}} \hat{\Delta}_{\mathbf{j}}^{\vphantom\dagger}}$ as a function of distance along the long direction $x$ of the bilayer system, for varying $J_{\perp}/t_{\parallel}$ and for widths $w=1,2,3,5$. We apply open boundary conditions in all directions. In the ladder systems ($w=1$), we find pronounced algebraic signals of pair-pair correlations throughout the whole system for all parameters, in line with previous findings presented in~\cite{qu2023}. When tuning the system towards the perturbative limit, pair-pair correlations are seen to converge towards spin-spin correlations $\braket{\hat{J}^{+}_{\mathbf{i}} \hat{J}^{-}_{\mathbf{j}}}$ of the mapped XXZ model (with $\hat{J}^{\pm}_{\mathbf{i}} = \hat{J}^{x}_{\mathbf{i}} \pm i \hat{J}^{y}_{\mathbf{i}}$), see the upper left panel of Fig.~\ref{fig:ppc}. Importantly, while the absolute values of pair-pair correlations rise for increasing $J_{\perp}/t_{\parallel}$, their corresponding decay exponent remains almost unchanged even down to $J_{\perp}/t_{\parallel} \sim 1$, which is the relevant regime for LNO~\cite{Luo2023}. In particular, fitted Luttinger exponents $K_{sc}$ (with $\braket{\hat{\Delta}^{\dagger}_{\mathbf{i}} \hat{\Delta}_{\mathbf{j}}^{\vphantom\dagger}} \propto |\mathbf{i} - \mathbf{j}|^{-K_{sc}}$) are $K_{sc} = 1.211(18)$ for $J_{\perp}/t_{\parallel} = 1$ and $K_{sc} = 0.946(3)$ for $J_{\perp}/t_{\parallel} = 20$. Similarly to the ladders, algebraic decay is observed for $w=3$ throughout the range of $J_{\perp}/t_{\parallel}$, cf. Fig.~\ref{fig:ppc}~(c). Here, the fitted Luttinger exponents are given by $K_{sc} = 0.82(2)$ for $J_{\perp}/t_{\parallel} = 5$ and $K_{sc} = 0.89(1)$ for $J_{\perp}/t_{\parallel} = 20$ For $w = 5$, we show results for $\delta = 0.5$ and in the perturbative regime $J_{\perp}/t_{\parallel} = 20$ for varying bond dimensions in Fig.~\ref{fig:ppc}~(e). Though variations of pair-pair correlations for increasing bond dimension are visible, an extrapolation to $\chi \rightarrow \infty$ matches the prediction from the XXZ model, suggesting long-range pairing order also away from $J_{\perp}/t_{\parallel} \gg 1$. 

In stark contrast to systems of odd widths, for $w=2$ at $\delta = 0.5$ we find at distances $j_x - i_x \gtrsim 10$ exponential behavior of pair-pair correlations, which, notably, has not been mentioned in previous numerical studies of the mixD bilayer $t-J$ model~\cite{qu2023}. 
A comparison with the perturbative XXZ model turns out as a useful tool to understand the origin of exponentially decaying pair correlations: In SU(2) symmetric Heisenberg ladders of even width and at zero magnetization, the formation of rung-singlets opens a spin-gap $\Delta_s$, which in turn leads to an exponential suppression of spin-spin correlations. In contrast, odd-width ladders have a vanishing spin-gap, and long-range correlations are observed~\cite{Troyer1994, White1994, Greven1996}. Similarly, we argue that the exponential decay of pair-pair correlations (even away from the perturbative limit) is an artifact of finite-size effects along the $y$-direction, driven by a finite charge pair gap $\Delta_{\text{pair}} = E(N) - E(N + 2) + J_{\perp}$. Here, $E(N)$ is the ground state energy at $\delta = 0.5$, i.e., with a total particle number of $N = l \times w$, and $E(N+2)$ corresponds to the energy of the system with one more particle in each layer compared to $\delta = 0.5$. We further add $J_{\perp}$ in $\Delta_{\text{pair}}$ to account for contributions from the Zeeman field in the effective XXZ description~\cite{supp_mat}. 

Indeed, the charge pair gap is seen to be finite throughout the whole parameter regime for $w=2$ at $\delta = 0.5$, as illustrated in Fig.~\ref{fig:ppc}~(f). Particularly, $\Delta_{\text{pair}}$ falls below the singlet-triplet spin gap in the Heisenberg model (where $\Delta_{s} \propto K$) for increasing $t_{\parallel}/J_{\perp}$, signaling a weaker exponential decay of pair-pair correlations when tuning the model away from the tightly-bound limit, matching observations in Fig.~\ref{fig:ppc}~(b). 

Away from $\delta = 0.5$, a finite magnetization in the effective model in the perturbative limit prevents the formation of a spin-singlet state, which in turn results in algebraic decay of spin-spin correlations even for finite, even-width systems. Likewise, pair-pair correlations in the bilayer model are seen to decay algebraically for $\delta \neq 0.5$, as shown for $\delta = 0.44$ in Fig.~\ref{fig:ppc}~(d). We note that in LNO, the coexistence of a strongly correlated state and a hole pocket in the $d_{z^2}$ band has been proposed to lead to self-doping between the $d_{z^2}$ and $d_{x^2 - y^2}$ orbitals, which is likely to slightly shift the doping in the $d_{x^2 - y^2}$ away from $\delta = 0.5$~\cite{cao2023flat, oh2023type}. In this case, the appearance of long-range pair-pair correlations is expected for all system widths in the single-band description. 

From our considerations, we conclude that in the thermodynamic limit, the model (with $J_{\parallel}\neq 0$ to break the emergent $\rm{SU(2)}$ symmetry) features quasi long-range pairing correlations up to a critical temperature determined by the Berezinskii-Kosterlitz-Thouless (BKT) transition $T_{\text{BKT}}$ where phase coherence occurs. We stress the direct correspondence of the decay of correlations in the mixD bilayer and XXZ model: When the effective model in the perturbative limit features an emerging SU(2) symmetry and forms spin singlets, correlations are exponential in the mixD model even away from $J_{\perp}/t_{\parallel} \gg 1$; however, when a finite magnetization prevents the formation of spin-singlets, correlations decay algebraically throughout the whole range of $J_{\perp}/t_{\parallel}$. Furthermore, there is only an insignificant change of the decay of pair-pair correlations when leaving the perturbative limit towards experimentally relevant regimes of $J_{\perp}/t_{\parallel} \sim 1$~\cite{Luo2023}, strongly suggesting that the key pairing physics of superconductivity in LNO is described by the XXZ universality class of hard-core bosons constituted by $s$-wave singlet pairs. Such controlled limits that capture the essential physics are absent in the plain-vanilla 2D Fermi-Hubbard model, where the ground state (not to mention the finite temperature phase diagram) is still under active debate due to the intricate competition between various phases~\cite{LeBlanc2015, Zheng2017, Jiang2019_science, Qin_absence_SC, MultiMess2021, Jiang_Kivelson, Arovas2022_rev, Qin_perspective, xu2023coexistence}. \\
 
\textbf{Binding energies and critical temperatures.---} Though long-range pair-pair correlations are necessary for any superconductor, their presence does not give any further insights into the nature and structure of the ground state. For this purpose, we compute inter-layer binding energies at $\delta = 0.5$ by evaluating $E_b = 2 E(N + 1) - E(N) - E(N + 2)$, and compare them to the spin gap $\Delta_s = E(N; S^z_{\text{tot}} = 1) - E(N; S^z_{\text{tot}} = 0)$.

\begin{figure}
\centering
\includegraphics[width=0.95\columnwidth]{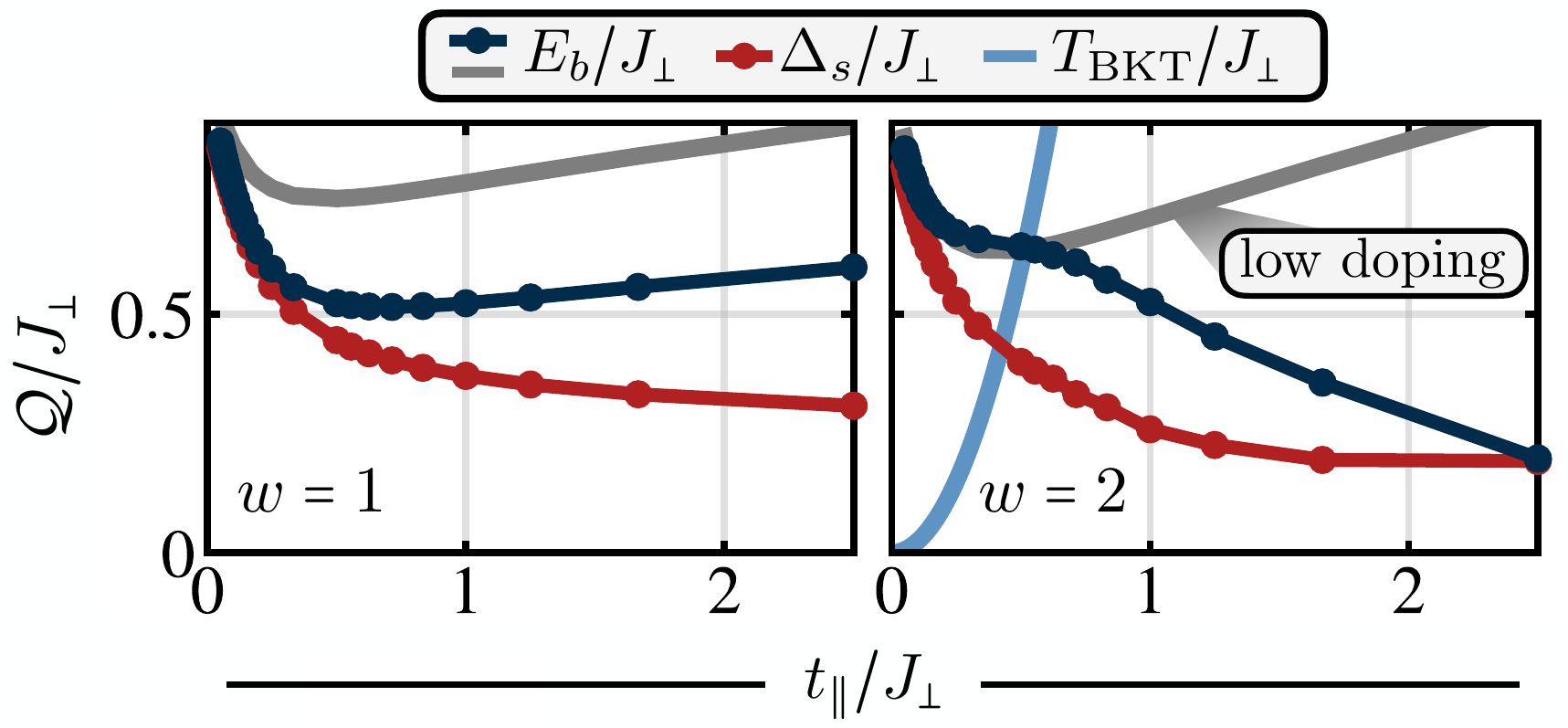}
\caption{\textbf{Binding energies.} Binding energies $E_b/J_{\perp}$ (dark blue) and spin gap $\Delta_s/J_{\perp}$ (red) as a function of $t_{\parallel}/J_{\perp}$ at $\delta = 0.5$ and $J_{\parallel} = 0$, for $w=1$ (left) and $w=2$ (right). For mixD ladders $w=1$, binding energies behave as predicted in the string picture~\cite{bohrdt2021strong}, where a large mobility of pairs for increasing $t_{\parallel}/J_{\perp}$ leads to enhanced binding energies. As a reference, binding energies in the dilute limit of a single hole pair are shown by grey solid lines. Meanwhile, due to the frustrating effect of moving charges, the spin gap decreases monotonously. For $w=2$, the binding energy in the zero doping limit features a similar structure as in the ladders. In contrast, large doping levels $\delta = 0.5$ permit the appearance of strongly overlapping chargon-chargon pairs, leading to a distinct drop of $E_b$ for $t_{\parallel}/J_{\perp} \gtrsim 0.6$. The crossing point of estimated critical temperatures of the BKT transition (corresponding to $w\rightarrow \infty$) in the perturbative regime (light blue line) with the binding energy coincides with the point of qualitative change of $E_b$, suggesting a BEC-BCS crossover as $t_{\parallel}/J_{\perp}$ is tuned.}
\label{fig:Eb}
\end{figure}

The left panel in Fig.~\ref{fig:Eb} shows results for mixD ladders, i.e. $w=1$. In the perturbative regime $t_{\parallel}/J_{\perp} \ll 1$, $E_b \approx \Delta_s \approx J_{\perp}$; each chargon-chargon as well as chargon-spinon pair is associated with energy $J_{\perp}$, while breaking up a singlet with fixed particle number also costs energy $J_{\perp}$. Away from the tightly-bound limit, the spin gap monotonously decreases, as growing sizes of the chargon-chargon bound states induce increasing frustration in the spin background~\cite{Schloemer2022_recon}. 
However, similar to the case of zero doping~\cite{bohrdt2021strong}, where we compute $E_b = 2 E(1) - E(0) - E(2)$ for a single hole pair (see grey solid lines in Fig.~\ref{fig:Eb}), the binding energy is observed to have a minimum around $t_{\parallel}/J_{\perp} \approx 0.5$, after which it starts to increase for further rising $t_{\parallel}/J_{\perp}$. In the low doping limit, it has been shown by some of us that this behavior is accurately captured within the string picture~\cite{Grusdt_strings, Bohrdt2020_partons, GrusdtX} of the mesonic bound states: an increasing mobility of the dopants leads to a significant kinetic contribution to the binding energy, resulting in an asymptotic scaling $E_b/J_{\perp} \sim (t_{\parallel}/J_{\perp})^{1/3}$~\cite{bohrdt2021strong}. Away from the perturbative limit, the monotonously descreasing spin-gap hence falls below the binding energy. Our numerical results demonstrate that the phenomenology is the same even at high doping $\delta = 0.5$ of the mixD ladders, though binding energies are renormalized to smaller values due to doping~\cite{supp_mat}.

When considering bilayer systems with widths $w>1$, the structure of the pairs fundamentally changes. In contrast to $w=1$ ladders, where the chargon-chargon bound states may overlap without destroying their confining strings, adding a second dimension allows for string-breaking processes~\cite{supp_mat}. In particular, this effect is expected to strongly influence the physics when the size of the bound pairs becomes comparable to the inter-pair distance for a given doping. Results for the binding energy and spin gap are shown for $w=2$ in the right panel of Fig.~\ref{fig:Eb}. In the dilute limit with only two holes (grey line), the binding energy is observed to feature a string-like behavior as expected. 
Likewise, in the perturbative regime $t_{\parallel}\ll J_{\perp}$ at doping $\delta = 0.5$, we find binding energies following the string prediction since string lengths $d \lesssim 1$ remain small compared to the average distance between hole pairs. Strikingly, this behavior extends well beyond the perturbative limit, where binding energies at $\delta = 0.5$ are seen to match predictions in the dilute limit up to $t_{\parallel}/J_{\perp}\approx 0.6$. However, for $t_{\parallel}/J_{\perp}\gtrsim 0.6$ -- where $d \gtrsim 1$~\cite{bohrdt2021strong} -- the binding energy starts to decrease for growing $t_{\parallel}/J_{\perp}$, approximately approaching the spin gap for large $t_{\parallel}/J_{\perp}$, which is expected in a BCS-like state. This, in turn, suggests the appearance of a BCS phase beyond $t_{\parallel}/J_{\perp} \gtrsim 0.6$, consisting of spatially extended pairs of holes.  

We further corroborate the appearance of a BEC-BCS crossover by estimating critical temperatures of the BKT phase ordering transition in the perturbative limit. In the 2D XXZ model with coupling $K$, extensive quantum Monte Carlo studies have quantified the phase transition, finding $T_{\text{BKT}}/K \approx 0.7$ ($0.6$) for $\Delta = 0$ ($0.95$)~\cite{Carrasquilla_2013, haldar2022study}. We estimate critical temperatures in the mixD bilayer model by assuming a small in-plane superexchange coupling $J_{\parallel}$, leading to an anisotropy close to the Heisenberg point in the effective XXZ description, $\Delta \lesssim 1.0$. Hence, following Eq.~\eqref{eq:XXZ} and assuming $\Delta = 0.95$, the BKT transition temperature is estimated by $T_{\text{BKT}}/J_{\perp} \approx 2.4 \left( t_{\parallel}/J_{\perp} \right)^2$, shown by the blue solid line in the right panel of Fig.~\ref{fig:Eb}. Indeed, we find that the critical temperature $T_{\text{BKT}}$ for phase coherence surpasses the binding energy at $t_{\parallel}/J_{\perp} \approx 0.6$, matching the point of qualitative change of $E_b$. Beyond this point, the superconducting transition is no longer driven by phase fluctuations and should be of BCS-type. In the BCS regime, the binding energy of a Cooper pair is given by $E_b = 2\Delta$ (with $\Delta$ the superconducting gap in the ground state), which implies critical temperatures of $T_c \sim 0.28 E_b$ for $t_{\parallel}/J_{\perp} \gtrsim 0.6$.

The resulting phase diagram of the mixD bilayer model is schematically shown in Fig.~\ref{fig:PT}~(a). In LNO, depending on the strength of the on-site Coulomb repulsion $U/t_{\parallel} \sim 5 - 10$, the predicted range of superexchange interactions is given by $t_{\parallel}/J_{\perp} \sim 0.7 -1.5$~\cite{Luo2023}, such that we predict the superconductor to be of BCS-type (though multi-band effects may renormalize the energy scales~\cite{oh2023type, yang2023strong}). We note that a complementary mean-field study of a related model found a similar BEC-BCS crossover phenomenology, however with quantitatively different results~\cite{lu2023superconductivity}. In order to experimentally verify the nature of the condensate, we propose to measure the specific heat of LNO under pressure, where a symmetric (asymmetric) shape is expected in a BEC-like (BCS-like) state as a function of the temperature~\cite{LORAM1994134, Wyk2016}. Measuring the shift of spectral weight of the optical conductivity across the superconducting phase transition may give additional insights into the nature of the condensate~\cite{lu2023superconductivity}. 

We speculate that the in-plane hopping $t_{\parallel}$ for systems of widths $w>1$ plays a similar role as nearest-neighbor particle repulsion in mixD ladders, where a related crossover from tightly bound pairs of holes (closed channel) at small repulsion to more spatially extended, correlated pairs of individual holes (open channel) at large repulsion has been proposed and studied in detail in~\cite{Lange2023_1, Lange2023_2, yang2023strong}. There, it was argued that the attraction of holes is ultimately mediated by the closed channel in analogy to a Feshbach resonance~\cite{Homeier2023}. Our simulations of extended systems similarly suggests Feshbash mediated pairing in bilayer nickelates, resulting in an effective attraction of spinon-chargon pairs due to the presence of the closed chargon-chargon channel. 

Regardless of whether the constituents of the superfluid are tightly bound chargon-chargon pairs (BEC) or overlapping Cooper pairs (BCS), binding energies of the order of the coupling $J_{\perp}$ suggest extraordinarily high critical temperatures in bilayer systems. Assuming an inter-layer coupling of $J_{\perp} \approx 0.3$ eV~\cite{Luo2023}, our results propose transition temperatures of the order of $T_c \approx 1000 $ K in the region of the BEC-BCS crossover, which is an order of magnitude larger than measured in LNO. We note however that the multi-band nature of LNO likely leads to strong suppressions of the condensation temperatures. For instance, a more sophisticated two-band model that takes into account Hund's coupling in a more rigorous manner has been shown to effectively reduce the coupling $J_{\perp}$ by a factor of four, which shifts the system deeper into the BCS phase and reduces its critical temperature~\cite{oh2023type}. Nevertheless, we stress that the physics in the perturbative limit (i.e. a description by an effective XXZ model) stays identical up to renormalization of the parameters, supporting the view that the single-band model captures the essential pairing physics. Considering both an effective reduction of $J_{\perp}$ due to the multi-band nature of LNO as well as the BCS prediction for $T_c$ leads to estimated critical temperatures of the mixD bilayer model that are indeed of the same order of magnitude as measured in LNO. Disorder effects may further suppress $T_c$ in bilayer nickelate materials, such that higher critical temperatures may be reached for cleaner samples. Though the above effects likely play a major role in determining the exact quantitative transition temperature of LNO, high $T_c$'s of the order of $J_{\perp}/2$ in the single-band, mixD bilayer model Hamiltonian near the crossover are very striking in their own right, and may open the path towards a more targeted design of materials possibly facilitating superconductivity above room temperature.

Lastly, we note that there exist intriguing similarities between the condensation of electron-hole pairs (excitons)~\cite{lozovik1976new, HANAMURA1977209} in bilayer semiconductors and superconductivity in bilayer nickelates. For example, high-temperature condensation of inter-layer excitons with large binding energies of $E_b \gtrsim  100$ meV has been demonstrated in bilayer transition metal dichalcogenide (TMD) semiconductors~\cite{Wang_exciton}. Additionally, a BEC-BCS crossover between tightly and weakly bound electron-hole pairs has been observed in bilayer quantum Hall systems by continuously tuning the pairing strength through variation of the layer separation~\cite{Liu_exciton}.

\textbf{Strange metallicity.---} Above the superconducting critical temperature of LNO, an extended region of strange metallicity with linear resistivity $\rho \propto T$ has been reported~\cite{Sun2023, zhang2023_zeroR}. Indeed, it has been shown that hard-core bosons on the 2D square lattice show a very similar behavior, with zero resistivity for $T < T_{\text{BKT}}$ and asymptotic linear resistivity $d\rho/dT \propto 1/\rho_s  $ above $T_{\text{BKT}}$, with $\rho_s$ the phase stiffness in the ground state ~\cite{Lindner2010}. We note that this is in stark contrast to weakly interacting Bose gases, where the resistivity saturates at high temperatures. This shows that, within the perturbative limit, the extended regime of linear in $T$ resistivity above the superconducting transition temperature as measured in LNO is captured in the effective model of tightly bound pairs, cf. Fig.~\ref{fig:PT}~(a).

We propose that, away from the perturbative limit but in vicinity of the conjectured BEC-BCS crossover, the behavior of the conductivity is nevertheless dominantly dictated by the conduction of pairs, conceivably leading to linear in $T$ resistivity in the bilayer system at experimentally relevant parameters $J_{\perp}/t_{\parallel} \sim 1$, cf. Fig.~\ref{fig:PT}~(a). Further studies of the mixD bilayer system are, however, necessary to pin down its properties away from the BEC-like limit. As numerical simulations are heavily limited in system size already in the ground state, transport simulations at finite temperature are beyond the reach of current state-of-the-art techniques. In contrast, with recent advances in ultracold atom quantum simulations~\cite{Bloch2008, Esslinger2010, Bohrdt2020, Gall2021Bilayer}, direct observation of the superconducting and potential strange metal phase in the bilayer mixD $t_{\parallel}-J_{\perp}-J_{\parallel}$ model is within reach, as we discuss in more detail in the following.

\textbf{Cold atom proposal.} Already before the discovery of superconductivity in LNO, the mixD bilayer model, Eq.~\eqref{eq:Hbl}, has been proposed to feature enhanced pairing and high superconducting critical temperatures in repulsively interacting strongly correlated systems~\cite{Dagatto1992, bohrdt2021strong}. Subsequently, real-space hole pairing has been observed experimentally in ultracold atom simulations in optical lattices by realizing the mixD setup through potential gradients~\cite{Hirthe2022} -- paving the way towards an experimental realization of a phase-coherent condensate in repulsive fermionic lattice systems with ultracold atoms.

In order to measure long-range pairing correlations using ultracold atom snapshots, we propose to hole dope the upper, while doublon doping the lower layer in a bilayer optical lattice~\cite{Lange2023_1, schlömer2024local}. After state preparation and freezing out in-plane tunneling in both planes, we propose to perform a global $\pi/2$ tunneling pulse between the two layers resonant with the transition from rung-singlets to interlayer doublon-hole pairs. This realizes a $\pi/2$ rotation within the subspace spanned by rung-singlets and doublon-hole pairs, while other transitions are either Pauli-blocked or off-resonant~\cite{supp_mat, schlömer2024local}. Measuring spatial correlations between doublon-hole pairs from snapshots then allows for a direct probe of pair-pair (superconducting) order with power-law decay for $T\leq T_{\text{BKT}}$, without the need for simultaneous spin-charge resolution. We note that in the plain-vanilla Fermi-Hubbard model, which has been in the spotlight of fermionic quantum simulators in recent years, the strong competition between different phases, low $T_c$'s or even absence of superconductivity in the ground state~\cite{Qin_absence_SC} renders an observation of long-ranged pairing order with ultracold atoms a real challenge. The mixD bilayer model, in turn, facilitates such an observation in state-of-the-art experiments owing to its large tunability and high predicted critical temperatures of the order of $J_{\perp}/2$. Furthermore, transport properties can be measured by relaxation of an imposed density
modulation~\cite{Brown2019badmetal}, enabling a direct observation of the strange metal and superconducting phases in the mixD $t_{\parallel}-J_{\perp}-J_{\parallel}$ model. This would allow for the simulation of 2D bilayer systems for a generic choice of Hamiltonian parameters, ultimately enabling realistic simulations of materials using analog quantum machines. A detailed experimental proposal can be found in Ref.~\cite{schlömer2024local}.

\textbf{Tuning pair correlations.---} The comparison to the perturbative limit of tightly bound inter-layer pairs $J_{\perp} \gg t_{\parallel}, J_{\parallel}$ further gives us an intuitive understanding of all appearing terms in the mixD $t_{\parallel}-J_{\perp}-J_{\parallel}$ Hamiltonian, Eq.~\eqref{eq:Hbl}. 
\begin{figure}
\centering
\includegraphics[width=0.82\columnwidth]{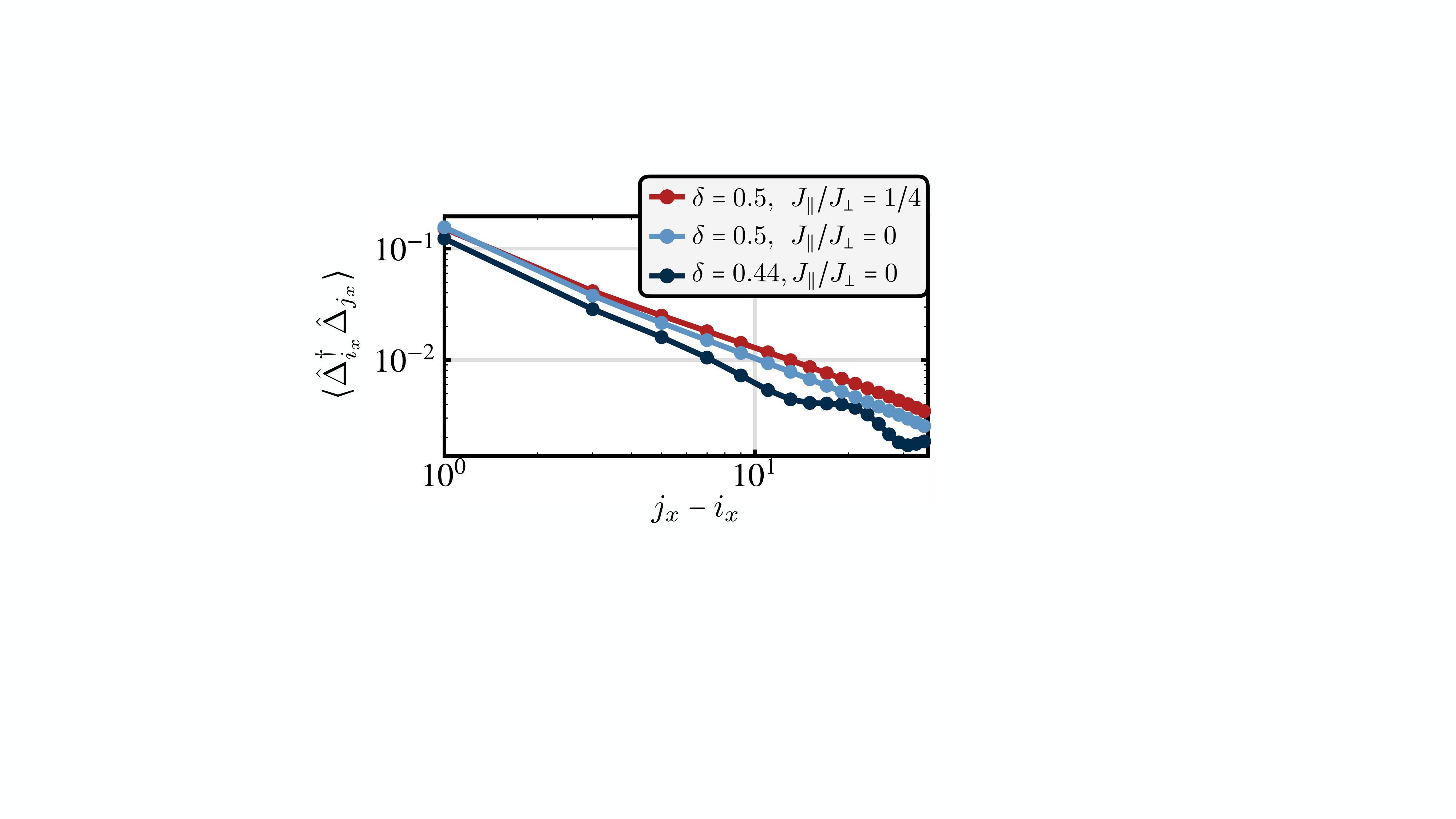}
\caption{\textbf{Tuning pair correlations.} Dependence of pair-pair correlations when tuning the filling $\delta$ and the ratio $J_{\parallel}/J_{\perp}$, for fixed $J_{\perp}/t_{\parallel} = 1$ (here shown for a system of size $w=1$, $l=64$). Finite in-plane AFM spin interactions $J_{\parallel}$ lead to an increase of pair-pair correlations. Doping the system away from $\delta = 0.5$ decreases pairing order and induces oscillatory boundary effects.}
\label{fig:params}
\end{figure}
Due to the in-plane magnetic interactions, an anisotropy $\Delta < 1$ is introduced in the effective XXZ model. This strengthens superconducting correlations compared to the isotropic case $\Delta = 1$: for small $\Delta$, the Luttinger decay exponent of correlations $\braket{\hat{J}^+_{i_x} \hat{J}^-_{j_x}}$ is proportional to $1+2\Delta/\pi$, i.e., smaller anisotropies $\Delta$ lead to slower power-law decay of correlations. Furthermore, doping the bosonic model away from $\delta = 0.5$ translates to finite magnetizations along $z$ in the XXZ model, leading to suppressed pair-pair correlations. We confirm these tendencies in DMRG simulations of the mixD bilayer model for experimentally relevant parameters, shown in Fig.~\ref{fig:params}. Note that the oscillatory behavior of correlations for $\delta = 0.44$ stem from Friedel modulations of the density that decay away from the open boundaries, and do not indicate charge order in the system.  

\textbf{Discussion.---} We have presented an extensive analysis of superconductivity in mixD bilayer systems by studying the single band $t_{\parallel}-J_{\perp}-J_{\parallel}$ model. By carefully analyzing finite size effects, we demonstrated that long-range pairing correlations emerge in the ground state, and quasi-long range power-law correlations below $T<T_{\text{BKT}}$, in the thermodynamic limit. We presented an analytically accessible limit of dominant inter-layer couplings, in which the model can be described by an effective spin-$1/2$ XXZ model. This allowed us to make predictions even away from this limit. Specifically, we proposed that the resistivity of the mixD bilayer system in the vicinity of the perturbative regime is dictated by the conduction of pairs, possibly explaining the linear in temperature resistivity measured in LNO above $T_c$.

Our study of binding energies at $\delta = 0.5$ proposes the appearance of a BEC-BCS crossover as the ratio $t_{\parallel}/J_{\perp}$ is tuned. This may lead to unexpected similarities with underdoped cuprates, where a similar Feshbach scenario has recently been proposed~\cite{Homeier2023}. With our understanding of all appearing terms in the mixD model Hamiltonian, we suggest to tune bilayer nickelates towards the BEC-BCS crossover point, e.g. through rare-earth substitutions as proposed in~\cite{pan2023effect}. Recent experiments suggest the appearance of superconductivity in trilayer nickelate compounds~\cite{zhang2024tri, Zhu2024_tri, Sakakibara_tri}. Performing a similar analysis for minimal models to trilayer systems and identifying relevant mechanisms may help to obtain a unified  understanding of nickelate superconductors~\cite{huo2024elec, LaBollita2024tri}.

By estimating transition temperatures of the phase ordering BKT transition, we predict unparalleled critical temperatures in the bilayer systems of the order of $J_{\perp}$ itself. This directly facilitates the preparation of a state with long-range pairing order in ultracold atom settings, and may pave the way towards the simulation, exploration and engineering of novel bilayer superconducting materials.

\textbf{Acknowledgments.---} We thank Tizian Blatz, Immanuel Bloch, Eugene Demler, Timon Hilker, Lukas Homeier, Hannah Lange, Matteo Mitrano, Subir Sachdev and  Ya-Hui Zhang for insightful discussions. This research was funded by the Deutsche Forschungsgemeinschaft (DFG, German Research Foundation) under Germany’s Excellence Strategy—EXC-2111—390814868 and by the European Research Council (ERC) under the European Union’s Horizon 2020 research and innovation programme (grant agreement number 948141 — ERC Starting Grant SimUcQuam). The data analyzed in the current study have been obtained using the SyTen package~\cite{hubig:_syten_toolk, hubig17:_symmet_protec_tensor_networ}.

\appendix
\widetext

\section{\underline{Supplementary Materials}}

\section{Strings in ladders and 2D systems}
\begin{figure}[!h]
\centering
\includegraphics[width=0.75\textwidth]{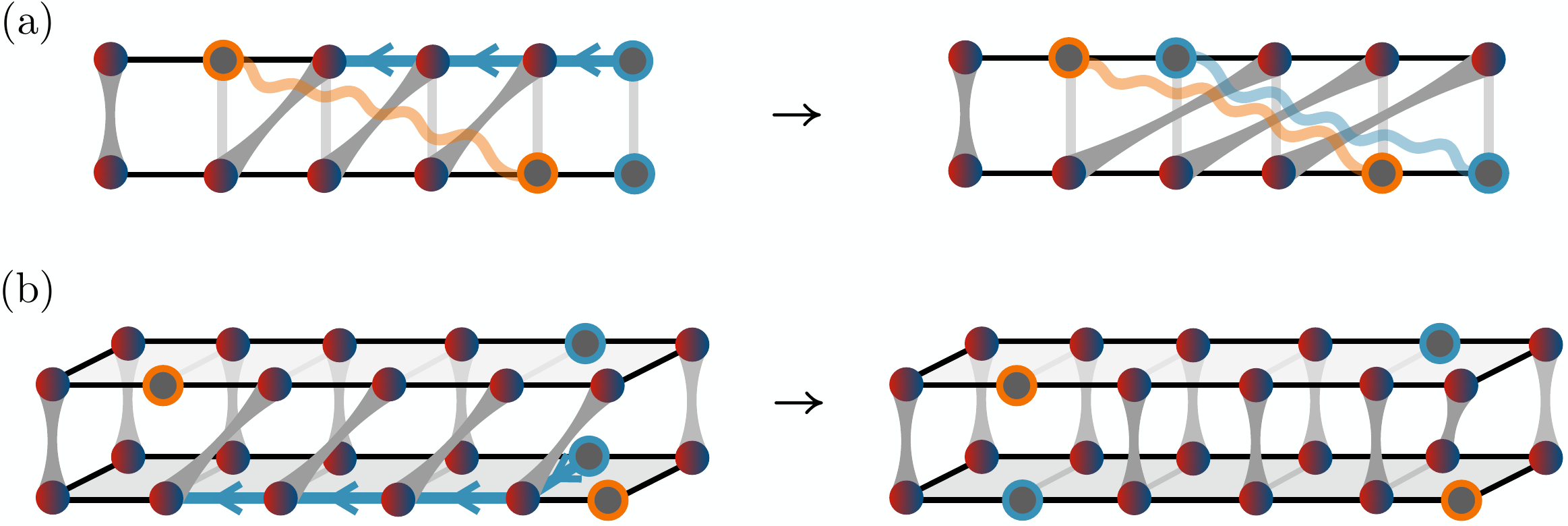}
\caption{\textbf{Re-traced strings.} An initially extended (orange) and localized (blue) chargon-chargon pair is shown both for a ladder, (a), and a system with width $w=2$, (b). In the case of the ladder, the upper blue hole can follow the orange string along the blue solid line, however the resulting strings (wiggly lines) do not cross and hence do not break. In extended systems along a second dimension, self-retracing paths are possible, leading to a fundamentally different structure and ultimately driving the BEC-BCS crossover discussed in the main text.}
\label{fig:strings}
\end{figure}

In the main text, we have argued that the possible structure of bound states is fundamentally different in ladders ($w=1$) compared to extended ladders ($w>1$) including 2D systems ($w\rightarrow \infty$) due to possible string-breaking processes, illustrated in Fig.~\ref{fig:strings}. In Fig.~\ref{fig:strings}~(a), two hole pairs are doped into the inter-layer singlet background for a ladder system, shown by the orange and blue circles. When the orange pair is separated, singlets are tilted in between the two holes, leading to the formation of a geometric string with linearly growing energy cost (orange wiggly line, left panel of Fig.~\ref{fig:strings}~(a)). When a chargon of a second pair (blue circles) follows the path shown by the blue line in the left panel in Fig.~\ref{fig:strings}~(a), the bound states can overlap, but strings are not allowed to cross and hence cannot break (Fig.~\ref{fig:strings}~(a), right panel). 

This structural restriction of strings appearing in ladders is lifted for $w>1$. Here, strings from separate chargon-chargon pairs can cross, as illustrated in Fig.~\ref{fig:strings}~(b). For instance, if the lower blue hole on the left-hand side in Fig.~\ref{fig:strings}~(b) follows the path shown by the blue line, the tilted singlets in between the extended orange pair can be re-traced, resulting in the state shown in the right panel of Fig.~\ref{fig:strings}~(b).

When the average size of pairs becomes comparable to the inter-pair distance, the string picture based on linearly growing string energies breaks down and has to be extended to include these processes. As shown in the main text, the binding energy deviates from the geometric string theory prediction for increasing $t_{\parallel}/J_{\perp}$ in systems with $w>1$, cf. Fig.~3~(b), leading to the BCS-BEC phenomenology that we predict in the mixD $t_{\parallel}-J_{\perp}-J_{\parallel}$ model. Whether the state in the BCS regime can be accurately described by extended, overlapping chargon-charcon pairs, or if a description in terms of spinon-chargon pairs is more appropriate will be discussed in future work.

\section{Numerical Details}
\begin{figure}[!h]
\centering
\includegraphics[width=0.75\textwidth]{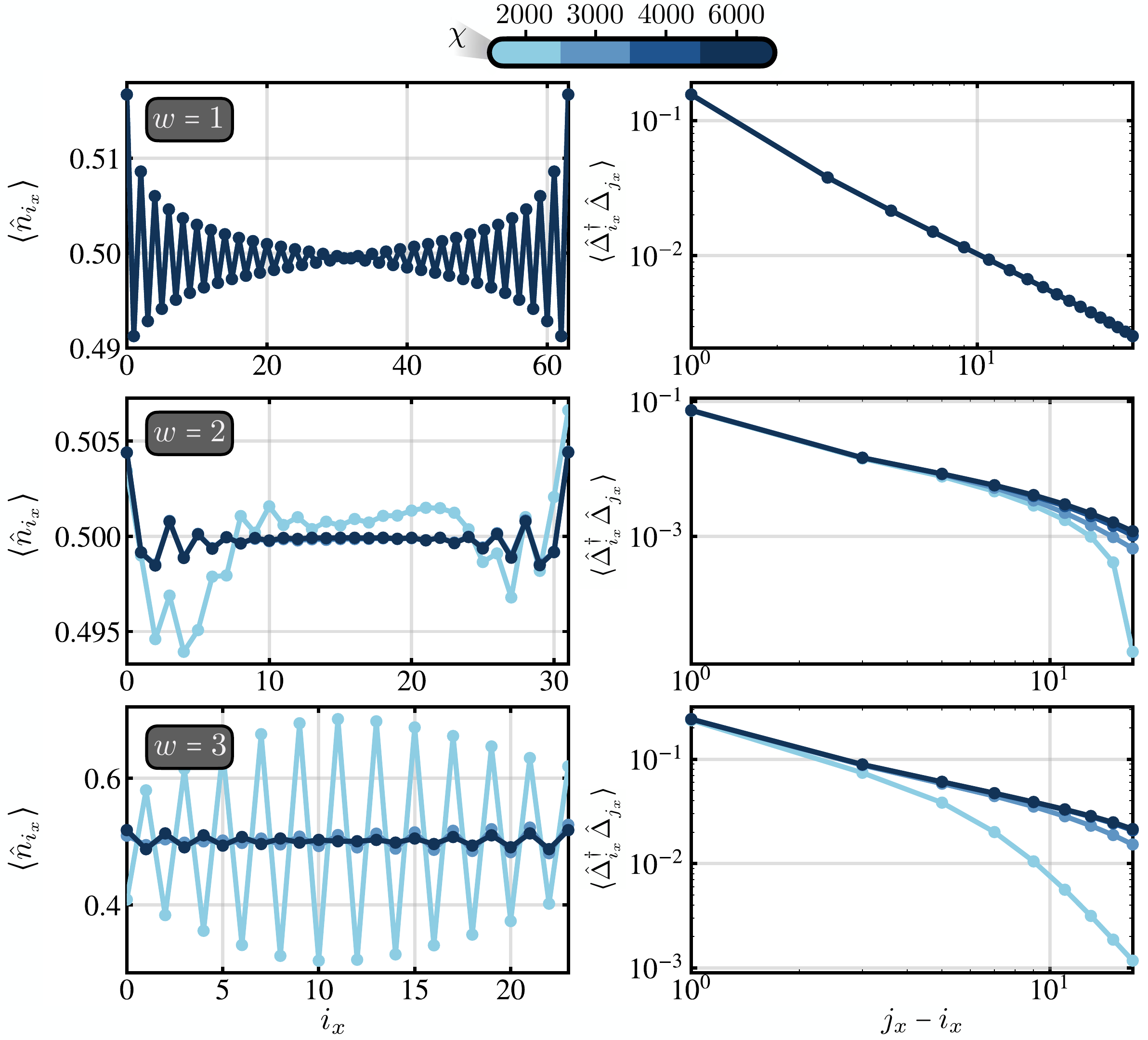}
\caption{\textbf{DMRG Convergence.} Convergence of the on-site particle density $\braket{\hat{n}_{i_x}}$ (left panels) as well as pair-pair correlations $\braket{\hat{\Delta}^{\dagger}_{i_x} \hat{\Delta}_{j_x}^{\vphantom\dagger}}$ (right panels) in the bilayer system for $w=1,2,3$. $t_{\parallel}/J_{\perp} = 1$ for $w=1,2$, and $t_{\parallel}/J_{\perp} = 10$ for $w=3$. In all cases, the particle densities as well as pair-pair correlations are observed to converge for $\chi > 4000$ ($\chi = 4000$ and $\chi = 6000$ are indistinguishable on the plot).}
\label{fig:conv}
\end{figure}

\underline{Convergence.} For our DMRG simulations, we use the SyTeN toolkit~\cite{hubig:_syten_toolk, hubig17:_symmet_protec_tensor_networ}. We explicitly implement the separate $\rm{U(1)}$ particle conservation symmetries in the two layers~\cite{Schloemer2022, Schloemer2022_recon}, resulting in a total $\rm{U(1)}^{\alpha = 1} \otimes \rm{U(1)}^{\alpha = 2}  \otimes \rm{U(1)}^{S^z_{\text{tot}}}$ symmetry of the system. We check convergence of the on-site particle density as well as pair-pair correlations in the bilayer systems. Fig.~\ref{fig:conv} shows $\braket{\hat{n}_{i_x}}$ as well as $\braket{\hat{\Delta}^{\dagger}_{i_x} \hat{\Delta}_{j_x}^{\vphantom\dagger}}$ for bond dimensions $\chi = 2000, \dots ,6000$. We see that both observables converge for all considered system widths ($\chi = 4000$ and $\chi = 6000$ are indistinguishable on the plot). \\

\begin{figure}
\centering
\includegraphics[width=0.55\textwidth]{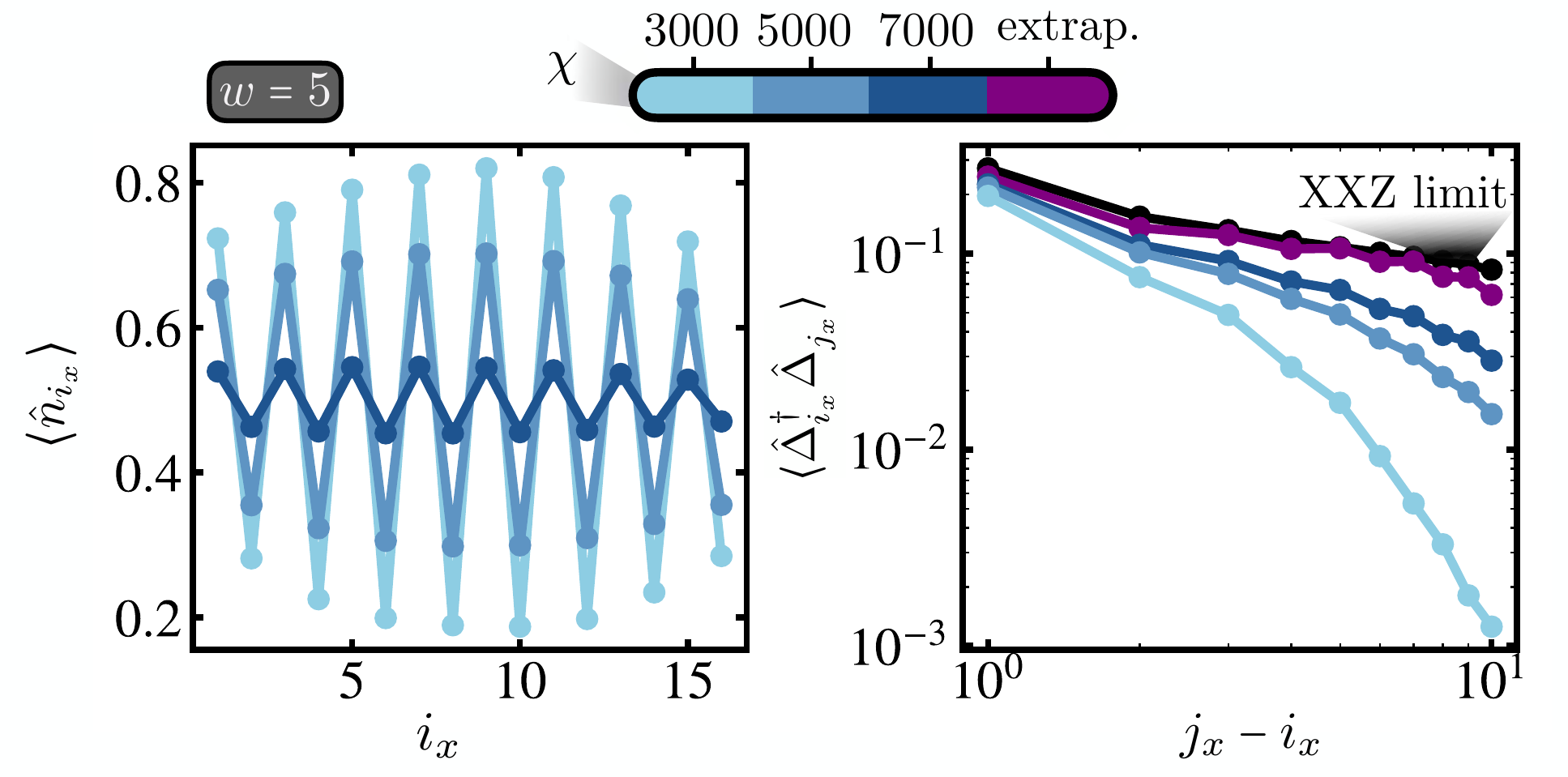}
\caption{\textbf{Wider bilayer systems.} On-site particle density and pair-pair correlations for a $L_x \times L_y = 16 \times 5$ system with $J_{\perp}/t_{\parallel} = 20$ and $\delta = 0.5$. Bond dimensions $\chi = 3000,5000,7000$ are shown by different shades of blue. The reference site for pair-pair correlations is chosen as $\mathbf{i} = [i_x = 3, i_y = 3]$. An extrapolation to infinite bond dimension is shown in purple. The black data corresponds to spin-spin correlations in the XXZ model.}
\label{fig:w5}
\end{figure}

\underline{Wider systems.} To make the bridge towards two-dimensional systems, we simulate a $L_x \times L_y = 16 \times 5$ system at doping $\delta = 0.5$ at various bond dimensions and close to the perturbative limit $J_{\perp}/t_{\parallel} = 20$, shown in Fig.~\ref{fig:w5}.
Both on-site density as well as pair-pair correlations show clear variations throughout the range of bond dimensions. We estimate the limit of $\chi \rightarrow \infty$ by a linear extrapolation of pair-pair correlations as a function of $1/\chi$, shown in purple in the right panel of Fig.~\ref{fig:w5}. While insufficient bond dimensions lead to a natural exponential decay of correlations, the extrapolation shows an algebraic signal that matches spin-spin correlations of the perturbative XXZ model, as discussed and expected from our considerations in the main text. While going significantly away from the perturbative regime is intractable for wide systems such as $w=5$ with current state-of-the-art techniques, the correspondence between the mixD bilayer $t-J_{\perp}-J_{\parallel}$ model and the XXZ model for $J_{\perp}/t_{\parallel} = 20$ and $w=5$ provides further evidence for the existence of long-range pairing order in the two-dimensional limit. \\

\underline{Binding energies.} In Fig.~\ref{fig:Eb} in the main text, we have presented binding energies $E_b/J_{\perp} = 2 E(N+1) - E(N) - E(N+2)$ of the mixD $t_{\parallel}-J_{\perp}-J_{\parallel}$ model at $\delta = 0.5$. With the implemented $\rm{U(1)}^{\alpha = 1} \otimes \rm{U(1)}^{\alpha = 2}  \otimes \rm{U(1)}^{S^z_{\text{tot}}}$ symmetry, $E(N+1)$ corresponds to a system with an additional particle in one layer compared to $\delta = 0.5$, while the other layer is doped with $N/2$ holes. We here compare resulting binding energies to a system where only the total number of particles is conserved in the bilayer system, i.e., with a $\rm{U(1)}^{\alpha = 1,2} \otimes \rm{U(1)}^{S^z_{\text{tot}}}$ symmetry. When simulating the mixD model with the latter (reduced) symmetry, we add a small perpendicular hopping $t_{\perp}/t_{\parallel} = 0.01$ between the planes, such that we calculate ground state energies of the Hamiltonian
\begin{equation}
\begin{aligned}
    \hat{\mathcal{H}} = -t_{\parallel} &\sum_{ \braket{\mathbf{i}, \mathbf{j}}, \sigma, \alpha} \hat{\mathcal{P}} \big(\hat{c}_{\mathbf{i}, \sigma, \alpha}^{\dagger} \hat{c}_{\mathbf{j}, \sigma, \alpha}^{\vphantom\dagger} + \text{h.c.} \big)\hat{\mathcal{P}} -t_{\perp} \sum_{\mathbf{i}, \sigma} \hat{\mathcal{P}} \big(\hat{c}_{\mathbf{i}, \sigma, 1}^{\dagger} \hat{c}_{\mathbf{i}, \sigma, 2}^{\vphantom\dagger} + \text{h.c.} \big)\hat{\mathcal{P}} \\ &+ J_{\parallel} \sum_{\braket{\mathbf{i}, \mathbf{j}}, \alpha} \left( \hat{\mathbf{S}}_{\mathbf{i},\alpha} \cdot \hat{\mathbf{S}}_{\mathbf{j},\alpha} - \frac{\hat{n}_{\mathbf{i},\alpha}\hat{n}_{\mathbf{j},\alpha}}{4} \right) + J_{\perp} \sum_{\mathbf{i}} \left( \hat{\mathbf{S}}_{\mathbf{i},1} \cdot \hat{\mathbf{S}}_{\mathbf{i},2} - \frac{\hat{n}_{\mathbf{i},1}\hat{n}_{\mathbf{i},2}}{4} \right).
\end{aligned}
\label{eq:SM_Hbl}
\end{equation}

\begin{figure}
\centering
\includegraphics[width=0.9\textwidth]{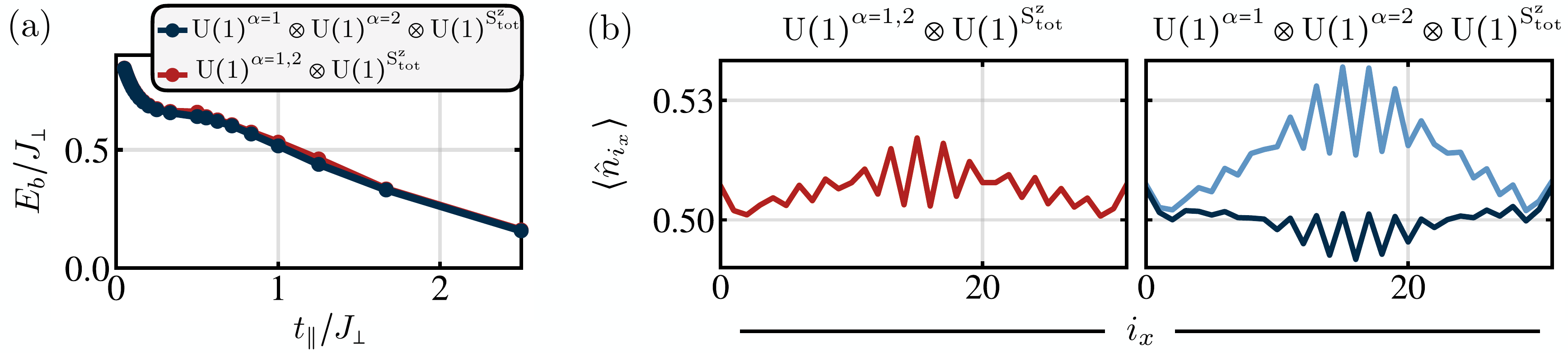}
\caption{\textbf{Different symmetries in DMRG calculations.} (a) Binding energies as a function of $t_{\parallel}/J_{\perp}$ when conserving the particle number in each layer (blue) compared to conserving only the total number of particles in the system (red). In the latter case, a small $t_{\perp}/t_{\parallel}$ is added to reach convergence. (b) Local densities $\braket{\hat{n}_{i_x}}$ for $J_{\perp}/t_{\parallel} = 1.6$ and particle number $N+1$. In the left panel, the total number of particles is conserved, such that the densities in the lower (dark red) and upper (light red - invisible) layers are identical. In the right panel, the particle number in each layer is conserved; the upper layer (light blue) has one hole less compared to $\delta = 0.5$, i.e., $N/2+1$ particles.}
\label{fig:Eb_SM}
\end{figure}

Results for the binding energies and particle densities are presented in Fig.~\ref{fig:Eb_SM}~(a) and (b), respectively, with $\chi = 6000$ ($\chi = 5000$) for the calculations with implemented $\rm{U(1)}^{\alpha = 1} \otimes \rm{U(1)}^{\alpha = 2}  \otimes \rm{U(1)}^{S^z_{\text{tot}}}$ ($\rm{U(1)}^{\alpha = 1,2} \otimes \rm{U(1)}^{S^z_{\text{tot}}}$) symmetry. While the density distributions are different due to the odd number of particles in the system, binding energies are almost indistinguishable.  \\

\underline{Perturbative regime.} In the main text, we have shown that the bilayer system is effectively described by a spin-$1/2$ XXZ model in the perturbative limit of dominating inter-layer spin couplings $J_{\perp}\gg t_{\parallel}, J_{\parallel}$, 
\begin{equation}
    \hat{\mathcal{H}}_{\text{XXZ}} =K \sum_{\braket{\mathbf{i}, \mathbf{j}}} \left( \hat{J}^x_{\mathbf{i}} \hat{J}^x_{\mathbf{j}} + \hat{J}^y_{\mathbf{i}} \hat{J}^y_{\mathbf{j}} + \Delta \hat{J}^z_{\mathbf{i}} \hat{J}^z_{\mathbf{j}} \right)  - J_{\perp} \sum_{\mathbf{i}} \hat{J}_{\mathbf{i}}^{z} - \frac{J_{\parallel}}{4} \sum_{\braket{\mathbf{i}, \mathbf{j}}} \left(\hat{J}^z_{\mathbf{i}} +  \hat{J}^z_{\mathbf{j}} \right) .
\label{eq:XXZ_SM}
\end{equation}
Here, we further compare the bilayer $t_{\parallel}-J_{\perp}-J_{\parallel}$ model to the effective spin description.
\begin{figure}
\centering
\includegraphics[width=0.9\textwidth]{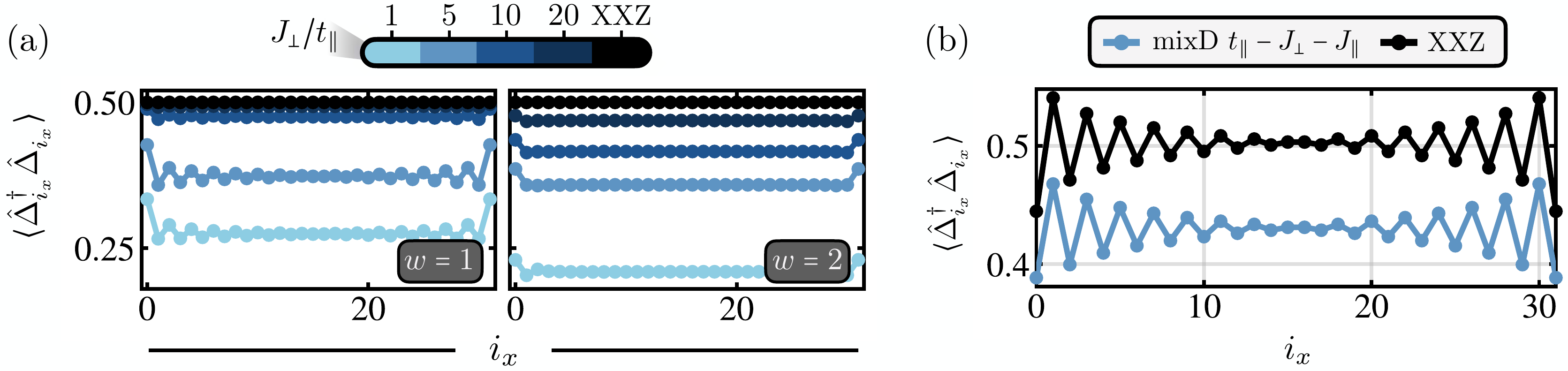}
\caption{\textbf{On-site pairs.} Number of on-site pairs $\braket{\hat{\Delta}_{i_x}^{\dagger} \hat{\Delta}_{i_x}^{\vphantom\dagger}}$ for varying $t_{\parallel}/J_{\perp}$ and for $w=1$ (left) and $w=2$ (right), both for $l=32$. In the perturbative regime, the number of pairs converges to the density $\delta = 0.5$, corresponding to $m + 1/2$ where $m$ is the magnetization of the mapped Heisenberg model. (b) $\braket{\hat{\Delta}_{i_x}^{\dagger} \hat{\Delta}_{i_x}^{\vphantom\dagger}}$ in the $t_{\parallel}-J_{\perp}-J_{\parallel}$ for $J_{\perp}/t_{\parallel} = 5$, $J_{\parallel}/t_{\parallel} = 0.3$ and for $w=1$. Oscillatory effects of the pair density in the vicinity of the boundary are captured within the effective XXZ model, and stem from the last term in Eq.~\eqref{eq:XXZ_SM} (black data).}
\label{fig:OSP}
\end{figure}
Fig.~\ref{fig:OSP}~(a) shows the local expectation values of on-site pairs, $\braket{\hat{\Delta}_{i_x}^{\dagger} \hat{\Delta}_{i_x}^{\vphantom\dagger}}$ for varying $J_{\perp}/t_{\parallel}$, $J_{\parallel} = 0$ and widths $w=1,2$, as well as $\braket{\hat{S}^z_{i_x}} + 0.5$ in the XXZ model. For $w=1$, the number of on-site pairs is seen to converge towards the Heisenberg limit more quickly than for $w=2$. This is in support of our speculations that the in-plane hopping $t_{\parallel}$ acts similarly as an inter-layer repulsive potential in mixD ladders~\cite{Lange2023_1, Lange2023_2} -- ultimately leading to the observation of a BEC-BCS crossover for $w>1$.  

In the main text, we primarily focused on the case $\delta = 0.5$, $J_{\parallel} = 0$, where the bilayer system can be reduced to the Heisenberg model in the perturbative limit. Here, we make further comparisons with the XXZ-like model away from this specific point.
We compute the on-site number of pairs for the mixD $t_{\parallel}-J_{\perp}-J_{\parallel}$ model for $l=32$, $w=1$, $J_{\perp}/t_{\parallel} = 5$, and $J_{\parallel}/t_{\parallel} = 0.3$, as shown in Fig.~\ref{fig:OSP}~(b). In the effective XXZ description, this leads to coupling parameters $K/t_{\parallel} = 0.8$ and an anisotropy of $\Delta = 0.8125$. On-site magnetizations $\braket{\hat{S}^z_{i_x}} + 0.5$ are shown in Fig.~\ref{fig:OSP}~(b) by black data points. The last term in Eq.~\eqref{eq:XXZ_SM} leads to oscillations of the magnetization in the vicinity of the boundary, visible also for the number of on-site pairs in the mixD $t_{\parallel}-J_{\perp}-J_{\parallel}$ model. We again stress that these effects are an artifact of open boundaries imposed by our DMRG simulations, and disappear in the thermodynamic limit. Nevertheless, the above further underlines our understanding of the appearing structure of pair densities in the $t_{\parallel}-J_{\perp}-J_{\parallel}$ by mapping to the effective XXZ spin-$1/2$ system.

We further study the effect of doping the system away from $\delta = 0.5$, 
\\

\underline{Pair charge gap.} When analyzing the decay of pair-pair correlations of the bilayer mixD model, we compared its pair charge gap $\Delta_{\text{pair}}$ to the spin gap $\Delta_s$ of the effective XXZ spin system, see Fig.~\ref{fig:ppc}~(d) in the main text. The latter is given by the singlet-triplet gap of the effective spin-$1/2$ model, i.e. $\Delta_s = E(S_{\text{tot}} = 1) - E(S_{\text{tot}} = 0)$. In the limit of vanishing in-plane spin interactions $J_{\parallel}$, Eq.~\eqref{eq:XXZ_SM} reduces to the Heisenberg model with an emergent $\rm{SU(2)}$ symmetry, where both the total spin $S_{\text{tot}}$ as well as the total magnetization along $z$, $S_{\text{tot}}^z$, commute with the Hamiltonian and are good quantum numbers. However, the effective Zeeman field $\propto J_{\perp} \sum_{\mathbf{i}}\hat{J}^z_{\mathbf{i}}$ in Eq.~\eqref{eq:XXZ_SM} splits the energy levels for total spin $S_{\text{tot}} = 1$. In order to calculate the genuine singlet-triplet spin gap, we explicitly remove the Zeeman contribution, and compute $\Delta_s = E(S^z_{\text{tot}} = 1) - E(S^z_{\text{tot}} = 1)$ of the Heisenberg model,
\begin{equation}
    \hat{\mathcal{H}}_{\text{Heis}} =K \sum_{\braket{\mathbf{i}, \mathbf{j}}} \left( \hat{J}^x_{\mathbf{i}} \hat{J}^x_{\mathbf{j}} + \hat{J}^y_{\mathbf{i}} \hat{J}^y_{\mathbf{j}} + \hat{J}^z_{\mathbf{i}} \hat{J}^z_{\mathbf{j}} \right),
\label{eq:XXZ_SM}
\end{equation}
where $\Delta_s/K = const$. Accordingly, to compare $\Delta_s$ to the pair charge gap in the bilayer model, we account for the above by computing $\Delta_{\text{pair}} = E(N) - E(N + 2) + J_{\perp}$, as shown in Fig.~2~(d).

\section{Experimental cold atom proposal}
In ultracold atom bilayer setups as realized in~\cite{Gall2021Bilayer, Hirthe2022}, the mixD $t_{\parallel}-J_{\perp}-J_{\parallel}$ model can be realized by simulating a bilayer Fermi-Hubbard (FH) model in the large $U/t$ limit (with $U$ the on-site repulsion) with a strong potential tilt $\Delta$ between the two planes~\cite{Trotzky, Dimitrova, Hirthe2022}. The potential offset $\Delta$ here effectively suppresses resonant tunneling between the layers, while virtual particle exchanges (and hence spin superexchange) remain intact. Specifically, if $\tilde{t}_{\parallel}$, $\tilde{t}_{\perp}$ denote in-plane and perpendicular hopping parameters in the simulated FH model, effective spin couplings in the limit $\tilde{t}_{\parallel}, \tilde{t}_{\perp} \ll U$ are given by~\cite{Duan2003, Trotzky, Hirthe2022} 
\begin{equation}
    J_{\perp} = \sum_{\pm} \frac{2\tilde{t}_{\perp}^2}{U \pm \Delta}, \quad J_{\parallel} = \frac{4 \tilde{t}_{\parallel}^2}{U}.
\end{equation}
Hence, the ratio $J_{\perp}/J_{\parallel}$ reads
\begin{equation}
    \frac{J_{\perp}}{J_{\parallel}} = \left( \frac{\tilde{t}_{\perp}}{\tilde{t}_{\parallel}} \right)^2 \frac{U^2}{U^2 - \Delta^2}.
\end{equation}
For $\tilde{t}_{\perp}<\Delta<U$, tunneling along the perpendicular direction in the effective $t-J$ model is suppressed, $t_{\perp} = 0$, while in-plane hopping remains unchanged, i.e., $t_{\parallel} = \tilde{t}_{\parallel}$. 

In order to measure pair-pair correlations in the mixD $t_{\parallel}-J_{\perp}-J_{\parallel}$, we propose a state preparation where one layer is doublon doped, while the other layer hole doped equivalently. To illustrate the protocol, we focus on the case where the energy-shifted layer by (positive) $\Delta$ is doublon doped; the other situation, which may be more natural to implement experimentally, follows in close analogy. The preparation gives rise to weak interactions $V$~\cite{Lange2023_2}, which can be tuned to values $V\ll J_{\perp}$ however, and can hence be neglected. Most importantly, the different doping of the two layers allows for coherent pair-creation processes $\sim \hat{\Delta}^{\dagger}_{\mathbf{j}}$ without changing the total fermion number. This can be brought to use for detecting long-range phase coherence:

\begin{figure}
\centering
\includegraphics[width=0.8\textwidth]{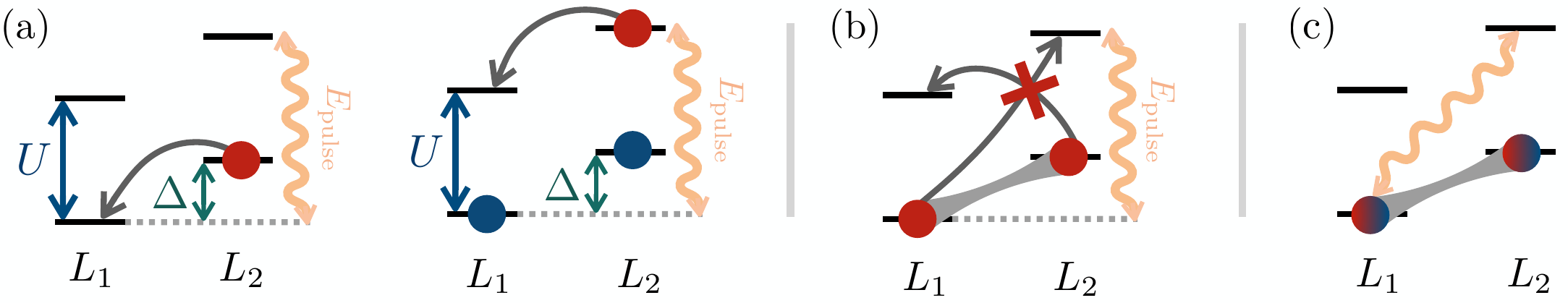}
\caption{\textbf{Measuring pair-pair correlations.} In a bilayer optical lattice, we suggest to measure coherent pair-pair correlations by doublon doping the layer with energy offset $\Delta$ (here denoted by $L_2$), while hole doping the other ($L_1$). Applying a $\pi/2$ pulse with energy $E_{\text{pulse}} = U + \Delta$ leads to the following effect depending on the local configuration: (a) Inter-layer hopping of single holes and doublons is off-resonant. (b) Transitions between local rung-triplets and doublon-hole pairs are Pauli-blocked. (c) Transitions between local rung-singlets and doublon-hole pairs are resonant, such that a $\pi/2$ tunneling pulse in the subspace spanned by rung-singlets and doublon-hole pairs (with the doublons in $L_2$) can be realized. Transitions to doublon-hole states with the doublons in $L_1$ are off-resonant.}
\label{fig:pulse}
\end{figure}

We start by considering the perturbative limit of the effective XXZ model, where in-plane correlations $\braket{\hat{J}^x_{\mathbf{i}} \hat{J}^x_{\mathbf{j}}}$ can be measured by applying a global $\pi/2$ basis rotation pulse and consecutively measuring in the $z-$basis, where $\braket{\hat{J}^z_{\mathbf{i}} \hat{J}^z_{\mathbf{j}}}$ correlations correspond to doublon-hole correlations. In terms of the mixD bilayer model, this $\pi/2$ rotation maps to a global $\pi/2$ tunneling pulse resonant with the transition from rung-singlets to inter-layer doublon-hole pairs. Corresponding energies for a resonant transition are given by $E_{\text{pulse}} = U + \Delta$ if the energy-shifted layer is doublon doped.

Away from the perturbative limit, though the Hilbert space is not spanned solely by rung-singlets and doublon-hole pairs, the above mentioned protocol nevertheless allows to measure pair-pair correlations, as all other transitions are either Pauli-blocked or off-resonant: 
\begin{enumerate}[(i)]
\item Inter-layer hopping of single holes and doublons are off-resonant, see Fig.~\ref{fig:pulse}~(a)
\item Transitions from local rung-triplet states to doublon-hole pairs are Pauli-blocked, see Fig~\ref{fig:pulse}~(b).
\item Transitions between rung-singlets and doublon-hole states (with doublons on the energy-shifted layer) are resonant, see Fig.~\ref{fig:pulse}~(c). This realizes a $\pi/2$ rotation within the subspace spanned by singlets and doublon-hole pairs. 
\end{enumerate}
Measuring spatial correlations between doublon-hole (dh) pairs, $C(\mathbf{i} - \mathbf{j}) = \braket{\hat{n}^{\text{dh}}_{\mathbf{i}} \hat{n}^{\text{dh}}_{\mathbf{j}}}$, then allows for a direct probe of superconducting (pair-pair) order, with power-law decay at temperatures below the 2D BKT transition $T\leq T_{\text{BKT}}$. In particular, simultaneous spin-charge resolution is no necessity for the protocol; density resolved snapshots are sufficient to measure coherent pair-pair correlations. However, with simultaneous spin-charge resolution, an additional rotation between rung-singlets and rung-triplets allows for the measurement of singlet, triplet, and doublon-hole densities separately, after which both singlet-singlet as well as dh-dh correlations can be evaluated to reveal the microscopic structure of the pairs in more detail.

\twocolumngrid
\bibliography{bilayer}

\end{document}